\documentclass[12pt]{article}
\usepackage{a4wide,amsmath,amssymb,alltt}

\newcommand\ie{i.e.\ }
\newcommand\eg{e.g.\ }

\newcommand\realcplx[1]{\texttt{#1}\{\texttt{Real},\texttt{Complex}\}}
\newcommand\lbrac{\symbol{123}}
\newcommand\rbrac{\symbol{125}}
\newcommand\grey[1]{\special{ps: .5 setgray}#1\special{ps: 0 setgray}}

\makeatletter
\def\reportno#1{\gdef\@reportno{#1}}
\def\@maketitle{%
  \hfill{\small\begin{tabular}[t]{r}%
    \@reportno
  \end{tabular}\par}%
  \vskip 2em%
  \begin{center}%
    \let \footnote \thanks
    {\large \@title \par}%
    \vskip 1.5em%
    {
      \lineskip .5em%
      \begin{tabular}[t]{c}%
        \@author  
      \end{tabular}\par}%
    \vskip 1em%
    {
     \@date}%
  \end{center}%
  \par
  \vskip 1.5em}
\makeatother

\begin{document}

\reportno{MPP--2006--159\\hep--ph/0611273}

\title{A Mathematica Interface for FormCalc-generated Code}

\author{T. Hahn \\
Max-Planck-Institut f\"ur Physik \\
F\"ohringer Ring 6, D--80805 Munich, Germany}

\date{November 21, 2006}

\maketitle

\begin{abstract}
This note describes a Mathematica interface for Fortran code generated 
by FormCalc.  The interfacing code is set up automatically so that 
only minuscule changes in the driver files are required.  The interface 
makes a function to compute the cross-section or decay rate available in 
Mathematica.  This function depends on the model parameters chosen for 
interfacing in the Fortran code.
\end{abstract}


\section{Introduction}

The FeynArts/FormCalc system \cite{FA,FC} is a fairly automated system
for generating Fortran code to numerically evaluate a wide class of
scattering and decay processes.  Thus far, this Fortran code was a
stand-alone, command-line driven application.  Many users added extra
functionality, either to the existing code, or by providing their own
frontend code \cite{HadCalc}.

The new Mathematica interface described herein turns the stand-alone
Fortran code into a Mathematica function for evaluating the
cross-section or decay rate as a function of user-selected model
parameters.  The benefits of such a function are obvious, as the whole
instrumentarium of Mathematica commands can be applied to them.  For
example, it is quite straightforward, using Mathematica's
\texttt{FindMinimum}, to determine the minimum (or maximum) of the
cross-section over a piece of parameter space.

Interfacing is done using the MathLink protocol.  The changes necessary
to produce a MathLink executable are quite superficial and affect only
the file \texttt{run.F}, where the user has to choose which model
parameters are interfaced from Mathematica.

To make the obvious even clearer, the cross-section is \emph{not}
evaluated in Mathematica, but in Fortran, and only the numerical results
computed by the Fortran code are transferred back to Mathematica.  One
thing one cannot do thus is to increase the numerical precision of the
calculation using Mathematica commands like \texttt{SetPrecision}.

The Fortran results are delivered to and accessed in Mathematica in a
way very similar to the \texttt{ReadData} utility which has been part of
the FormCalc distribution for quite a while already.  The interfacing
code for Mathematica is included from FormCalc Version 5.1 onward.

\section{Setting up the Interface}

The model parameters are specified in the file \texttt{run.F}.  Typical
definitions for stand-alone code look like (here from an MSSM 
calculation):
\begin{verbatim}
   #define LOOP1 do 1 TB = 5, 50, 5
   #define LOOP2 MA0 = 500
   ...
\end{verbatim}
These lines declare \texttt{TB} ($= \tan\beta$) to be scanned from 5 to
50 in steps of 5 and set \texttt{MA0} ($= M_{A^0}$) to 500 GeV.  To be
able to specify \texttt{TB} in Mathematica instead, the only change is
\begin{verbatim}
   #define LOOP1 call MmaGetReal(TB)
\end{verbatim}
Such invocations of \texttt{MmaGetReal} and its companion subroutine
\texttt{MmaGetComplex} serve two purposes.  At compile time they
determine with which arguments the Mathematica function is generated 
(for details see below), and at run time they actually transfer the 
function's arguments to the specified Fortran variables.
\begin{itemize}
\item\texttt{call MmaGetReal($r$)} \\
read the real parameter $r$ from Mathematica,
\item\texttt{call MmaGetComplex($c$)} \\
read the complex parameter $c$ from Mathematica.
\end{itemize}
Note that without a separate \texttt{MmaGetReal} call, \texttt{MA0} 
would still be fixed by the Fortran statement, \ie not be accessible 
from Mathematica.

Once the makefile detects the presence of these subroutines, it
automatically generates interfacing code and compiles a MathLink
executable.  For a file \texttt{run.F} the corresponding MathLink
executable is also called \texttt{run}, as in the stand-alone case. 
This file is not started from the command-line, but used in Mathematica 
as
\begin{verbatim}
   Install["run"]
\end{verbatim}

\section{The Interface Function in Mathematica}

After loading the MathLink executable with \texttt{Install}, a
Mathematica function of the same name is available.  For definiteness,
we will call this function `\texttt{run}' in the following since 
`\texttt{run.F}' is the default parameter file.  This function has the 
arguments (grey parts are optional)
\begin{alltt}
   run[sqrtS, arg1, arg2, ..., options]

   run[\lbrac{}sqrtSfrom, sqrtSto\grey{, sqrtSstep}\rbrac{}, arg1, arg2, ..., options]
\end{alltt}
The first form computes a differential cross-section at $\sqrt s = 
\mathtt{sqrtS}$.  The second form computes a total cross-section for 
energies $\sqrt s$ varying from \texttt{sqrtSfrom} to \texttt{sqrtSto} 
in steps of \texttt{sqrtSstep}.  This is in one-to-one correspondence 
with the command-line invocation of the stand-alone executable.

The \texttt{arg1}, \texttt{arg2}, \dots, are the model parameters
declared automatically by the presence of their \realcplx{MmaGet} calls
(see above).  They appear in the argument list in the same order as the
corresponding \realcplx{MmaGet} calls.

Other parameters are specified through the \texttt{options} (default 
values are written on the right-hand sides):
\begin{itemize}
\item\verb|Polarizations -> "UUUU"|,
the polarizations of the external particles, specified as in the 
stand-alone version, \ie a string of characters for each external leg:
\begin{center}
\begin{tabular}{|cl|cl|} \hline
\texttt{U} & unpolarized, &
	\texttt{L} & longitudinal polarization, \\
\texttt{T} & transversely polarized, \quad &
	\texttt{-} & left-handed polarization, \\
&&	\texttt{+} & right-handed polarization. \\ \hline
\end{tabular}
\end{center}

\item\verb|Serial -> {}|,
a range of serial numbers, specified as \texttt{\lbrac
serialfrom\grey{, serialto, serialstep}\rbrac} (grey parts optional).  
The concept of serial numbers, used to parallelize parameter scans, is
described in Ref.~\cite{LL04}.  This option applies only to parameters 
scanned by Fortran do-loops in the parameter statements.  Parameters 
read from Mathematica are unaffected by this option.

\item\verb|SetNumber -> 1|,
a set number beginning with which parameters and data are stored 
(see next Section).

\item\verb|ParaHead -> Para|,
the head under which parameters are stored, \ie parameters are 
retrievable from \verb|parahead[setnumber]| (see next Section).

\item\verb|DataHead -> Data|,
the head for the data storage, \ie data are retrievable from 
\verb|datahead[setnumber]| (see next Section).

\item\verb|LogFile -> ""|,
a log-file to save screen output in.  An empty string indicates no 
output redirection, \ie the output will appear on screen.
\end{itemize}

\section{Return values, Storage of Data}

The return value of the generated function is an integer which records 
how many parameter and data sets were transferred.  Assigning parameter 
and data sets as the data become available has several advantages:
\begin{itemize}
\item
the return value of \texttt{run} is an integer rather than a large, 
bulky list,

\item 
the parameters corresponding to a particular data set are easy to
identify, \eg \texttt{Para[4711]} contains the parameters corresponding
to \texttt{Data[4711]},

\item 
most importantly, if the calculation is prematurely aborted, the 
parameters and data transferred so far are still accessible.
\end{itemize}
Both, the starting set number and the heads of the parameter and data 
assignments can be chosen with the options \texttt{SetNumber}, 
\texttt{ParaHead}, and \texttt{DataHead}, respectively.

The parameters which are actually returned are chosen by the user in the
\texttt{PRINT$n$} statements in \texttt{run.F} in much the same way 
as parameters are selected for printout in the stand-alone code.  To 
specify that \texttt{TB} and \texttt{MA0} be returned, one needs the 
definitions
\begin{verbatim}
   #define PRINT1 call MmaPutReal("TB", TB)
   #define PRINT2 call MmaPutReal("MA0", MA0)
\end{verbatim}
Notwithstanding, parameters can still be printed out, in which case 
they end up in the log file (or on screen, if no log file is chosen).  
To transfer \eg \texttt{TB} to Mathematica \emph{and} print it out, one 
would use
\begin{verbatim}
   #define PRINT1 call MmaPutReal("TB", TB)
   #define PRINT2 SHOW "TB", TB
\end{verbatim}
An analogous subroutine exists of course for complex parameters, too:
\begin{itemize}
\item\texttt{call MmaPutReal($s$, $r$)} \\
transfer the real parameter $r$ to Mathematica under the name $s$,

\item\texttt{call MmaPutComplex($s$, $c$)} \\
transfer the complex parameter $c$ to Mathematica under the name $s$.
\end{itemize}
The parameters are stored in the form of rules in Mathematica, \ie as
\textit{name}\verb| -> |\textit{value}.  The first argument specifies
the left-hand side of this rule.  It need not be a symbol in the strict
sense, but can be an arbitrary Mathematica expression.  But note that in
particular the underscore has a special meaning in Mathematica and may
not be used in symbol names.  The second argument is then the right-hand
side of the rule and can be an arbitrary Fortran expression containing
model parameters, kinematic variables, etc.

The following example demonstrates the form of the parameter and data
assignments.  Shown are results of a differential cross-section for a
$2\to 2$ reaction at one point in MSSM parameter space.  Within the 
data the varied parameter is $\cos\theta$, the scattering angle.
\begin{verbatim}
   Para[1] = { TB -> 1.5, MUE -> -1000., MSusy -> 1000.,
               MA0 -> 700., M2 -> 100. }

   Data[1] = { DataRow[{500., -0.99},
                       {0.10592302458950732, 0.016577997941111422},
                       {0., 0.}],
               DataRow[{500., -0.33}, 
                       {0.16495552191438356, 0.014989931149150608},
                       {0., 0.}], 
               DataRow[{500., 0.33},
                       {0.2986891221231292, 0.015013326141014818},
                       {0., 0.}],
               DataRow[{500., 0.99},
                       {0.5071238252157443, 0.012260927614082411},
                       {0., 0.}] }
\end{verbatim}
The \texttt{DataRow[$v$,\,$r$,\,$e$]} function has three arguments:
\begin{itemize}
\item
the independent kinematic variables ($v = \{\sqrt s, \cos\theta\}$ above),

\item
the cross-section or decay-rate results ($r$ = \{tree-level result,
one-loop correction\} above), and

\item
the respective integration errors ($e = \{0, 0\}$ above, as this example
originates from the computation of a differential cross-section where no
integration is performed).
\end{itemize}

\section{Using the Generated Mathematica Function}

To the Mathematica novice it may not be obvious how to use the function 
described above to analyse data, produce plots, etc.

As an example, let us produce a contour plot of the cross-section in the
$M_{A^0}$--$\tan\beta$ plane.  It is assumed that the function
\texttt{run} has the two parameters \texttt{MA0} and \texttt{TB} in its
argument list:
\begin{verbatim}
   Install["run"]

   xs[sqrtS_, MA0_, TB_] := (
     run[{sqrtS, sqrtS}, MA0, TB];
     Data[1][[1,2]]
   )

   ContourPlot[xs[500, MA0, TB], {MA0, 100, 500}, {TB, 5, 50}]
\end{verbatim}
The function \texttt{xs} runs the Fortran code and selects the data to
plot.  The first argument of \texttt{run}, \verb|{sqrtS, sqrtS}|,
instructs the Fortran code to compute the total cross-section for just
one point in energy.  We then select the first (and only)
\texttt{DataRow} in the output and choose its second argument, the
cross-section results: \verb|Data[1][[1,2]]|.

This example can be extended a little to produce a one-dimensional
plot where \eg for each value of $\tan\beta$ the minimum and maximum
of the cross-section with respect to $M_{A^0}$ is recorded:
\begin{verbatim}
   << Graphics`FilledPlot`

   xsmin[sqrtS_, TB_] :=
     FindMinimum[xs[sqrtS, MA0, TB], {MA0, 100}][[1]]
   xsmax[sqrtS_, TB_] :=
     -FindMinimum[-xs[sqrtS, MA0, TB], {MA0, 100}][[1]]

   FilledPlot[{xsmin[500, TB], xsmax[500, TB]}, {TB, 5, 50}]
\end{verbatim}

\section{Summary}

The new Mathematica interface makes it easy to turn Fortran code
generated by FormCalc, thus far a stand-alone, command-line driven
application, into a Mathematica function for numerically evaluating the
cross-section.  The generated function depends on any kinematical or 
model parameter the user chooses.  The Fortran code is accessed via the 
MathLink protocol.

The advantages of such a function are obvious, as all of Mathematica's
sophisticated tools for visualization, numerical analysis, etc.\ can
immediately be applied to it.  Interfacing with Mathematica requires
only minuscule changes to the parameter file \texttt{run.F}.  The
necessary changes in the build process are taken care of by the
makefile.  The interfacing code is available from FormCalc Version 5.1
onward.

The author welcomes bug and performance reports, as well as suggestions
for improvements at hahn@feynarts.de.

\end{document}